\documentclass[conference]{IEEEtran}
\IEEEoverridecommandlockouts
\usepackage{graphicx}  
\usepackage{epstopdf}
\usepackage{color}
\usepackage{float} 
\usepackage{blindtext}
\usepackage{xfrac}
\usepackage{setspace}
\usepackage{amsmath,amssymb, amsthm}
\usepackage{multirow}
\usepackage[tight,footnotesize]{subfigure}
\usepackage[colorinlistoftodos, textwidth=3.4cm, shadow]{todonotes}
\usepackage{balance}
\usepackage{xcolor}

\usepackage{color}   
\usepackage{listings}
\usepackage[normalem]{ulem}
\newcommand\angelo[1]{\textcolor{blue}{#1}}

\newcommand\bart[1]{\textcolor{brown}{#1}}
\definecolor{dark-green}{HTML}{006600}

\begin{document}

\title{{Properties of the Tangle for uniform random and random walk tip selection}}

\author{
Bartosz Ku\'smierz\IEEEauthorrefmark{1}, William Sanders, Andreas Penzkofer, Angelo Capossele and Alon Gal\\
   IOTA Foundation, c/o Nextland Strassburgerstra{\ss}e 55,  10405 Berlin, Germany\\
    \IEEEauthorrefmark{1} Dept. of Theoretical Physics, Wroc{\l}aw University of Science and Technology,
    50-370 Wroc{\l}aw, Poland\\
    Email: \{bartosz.kusmierz, william.sanders, andreas.penzkofer, angelo.capossele, alon.gal\}@iota.org
}

\maketitle

\begin{abstract}
The growing number of applications for distributed ledger technologies is driving both industry and academia to solve the limitations of blockchain,  particularly its scalability issues. 
Recent distributed ledger technologies have replaced the blockchain linear structure with a more flexible directed acyclic graph in an attempt to accommodate a higher throughput. 
Despite the fast-growing diffusion of directed acyclic graph based distributed ledger technologies, researchers lack a basic understanding of their behavior.
In this paper we analyze the {\it Tangle}, a directed acyclic graph {that is used  (with certain modifications) in various protocols such as IOTA, Byteball, Avalanche or SPECTRE.}

Our contribution is threefold. 
%
%
First, we run simulations in a continuous-time model to examine tip count stability and cumulative weight evolution while varying the rate of incoming transactions. 
In particular we confirm analytical predictions on the number of tips with uniform random tip selection strategy.
Second, we show how different tip selection algorithms affect the growth of the Tangle. 
Moreover, we explain these differences by analyzing the spread of exit probabilities of random walks.  
Our findings confirm analytically derived predictions and provide novel insights on the different phases of growth of cumulative weight as well as on the average time difference for a transaction to receive its first approval when using distinct tip selection algorithms.
Lastly, we analyze simulation overhead and performance as a function of Tangle size and compare results for different tip selection algorithms. 
\end{abstract}

\section{Introduction}

Satoshi Nakamoto, Bitcoin's creator(s),  proposed a new decentralized payment system \cite{bit} based on a trustless peer-to-peer network. Bitcoin is essentially a protocol for reaching consensus on a chronologically ordered log of transactions called a {\it{blockchain}}. This data structure is now a cornerstone of many other Distributed Ledger Technologies, or  DLTs for short, \cite{ethereum, algorand, cardano, cryptocurrencySurvey}.
Blockchain based DLTs are finding applications in other fields, ranging from financial services such as digital assets, remittance, and online payments to smart contracts, public services, reputation systems, and data marketplaces~\cite{zheng2016}.

However, blockchains have inherent scalability limitations: since transactions, or {\it{blocks}}, are limited in size and frequency, blockchains have a limited throughput.  This issue hinders the adoption of cryptocurrencies in contexts such as the Internet of Things (IoT) \cite{dorri2017}.

To overcome scalability issues, several techniques have been proposed ranging from increasing block size and frequency to sidechains \cite{croman2016}, ``layer-two'' structures like Lightning Network \cite{light}, Sharding \cite{luu2016}, and different consensus mechanisms, \cite{algorand, cardano, casper}.

In an attempt to increase throughput, some recent DLT's have replaced the  blockchain with a {\it{DAG}}, a directed acyclic graph. This approach is used in IOTA \cite{I_WP} and other new protocols; see \cite{Lewenberg2015, byteball, Boyen2016, spectre, meshcash, conflux, dexon, vegvisir}.  DAG based protocols reach consensus on a {\it{partially}} ordered log of transactions, which in turn, allow the log to have ``width'' and increase the throughput of the system.  

In this paper, we focus on the DAG based IOTA protocol introduced in \cite{I_WP}, although our results may have wider applications. In this DLT, transactions are recorded in a DAG dubbed the {\it Tangle}. The vertices in the Tangle are transactions. If there is an edge between transactions $x\leftarrow y$, we say that $y$ approves $x$, and we say that $y$ indirectly approves $x$ if there is a directed path from $y$ to $x$.  A transaction with no approvers is called a {\it tip}.  Under the{\iffalse IOTA\fi} protocol discussed in \cite{I_WP}, all incoming transactions attach themselves to the Tangle by approving two (not necessarily distinct) tips.  

A critical aspect of the Tangle{\iffalse IOTA protocol\fi} is the algorithm used to select the tips which a new transaction will attach itself to. Several such algorithms are discussed in \cite{I_WP}.  The {\it Uniform Random Tip Selection}, or {\it URTS}, is the most simple algorithm: we simply  select a tip from the set of all available tips with a uniform random distribution.
%
Another algorithm is the Monte Carlo Markov Chain or MCMC: here we select the tip at the end of a random walk\footnote{We actually mean an antiwalk, i.e.\ a walk moving backwards along the arrows.} beginning at the first transaction in the Tangle. The random walk can be biased towards transactions with large cumulative weight \cite[Section 4.1]{I_WP} and the amount of bias is determined by a parameter $\alpha\ge 0$. When  $\alpha = 0$, we dub the MCMC as {\it Unbiased Random Walk} or {\it URW}. When  $\alpha > 0$, we dub the MCMC as {\it Biased Random Walk} or {\it BRW}.
In this article, we will primarily focus on the  URTS and  URW selection algorithms. However, we also discuss   BRW in section~\ref{bias}.

For a transaction $x$, we study the cumulative weight of $x$, which is\footnote{The definition in \cite[Section 2]{I_WP} is more complex and allows for some transactions to weighted by their proof of work. However, this simplified definition is more suitable for this paper and is also the one currently used in IOTA.} one plus the number of transactions indirectly approving $x$ \cite[Section 2]{I_WP}. We observe that the cumulative weight experiences two phases of growth, first exponential and then linear, confirming the predictions of \cite[Section 3.1]{I_WP}. We also study the number of tips in the Tangle and note that with both tip selection algorithms this number fluctuates around a constant value.  For various rates of incoming transactions, we record the average and standard deviation of the number of tips. We also analyze the time until a transaction receives its first approval, and we observe the average time is higher under URW. We also study the distributions of exit probabilities, which is the probability of selecting a particular tip given a specific tip selection algorithm.   Lastly, we discuss how the performance and computational overhead of our simulations depend on the tip selection mechanisms and compare the results of both.

The results give insight into the basic behavior of the Tangle and are important for developing and evaluating the performance and security of the Tangle based{\iffalse IOTA \fi} protocols.  For instance, the number of tips and time until first approval are related to computing the throughput of the system.  Also, the security of the system depends upon preventing attackers from manipulating the growth of the cumulative weight of a transaction after making a conflicting transaction.

The  other published works related to the Tangle focus on other areas.  Bramas \cite{bramas2018} analyses the security of the IOTA protocol,  and studies the confirmation level of a transaction $x$, i.e.\ the probability of selecting a tip which indirectly approves $x$. 
More recently, Zander et al.,~\cite{dagsim} propose a continuous time and multi-agent simulation framework for DAG-based cryptocurrencies. Their result focuses on the transaction attachment probabilities and shows that agents with low latency and high connection degree have a higher probability of having their transactions accepted in the network.  Recently, in  \cite{Bob} Ferraro et al.\ propose a new tip selection algorithm combining both URTS and BRW.  

The paper is organized as follows: In Section \ref{Tangle} we introduce the Tangle and discuss basic concepts such as cumulative weight and  tip selection mechanisms. We follow with Section \ref{results} where we present our results on cumulative weights, evolution of the number of tips, average time until the first confirmation and overhead of simulations. The presented data shows noticeable differences in Tangles grown employing either URTS or URW, which is explained through the differing shape of {\it exit probabilities}. In Section \ref{sec:safety}, we discusses tip selection algorithms and the safety of the network. Lastly, in Section   \ref{sec:disc_conslusion} we conclude our findings.  

\section{The Tangle}\label{Tangle}

\begin{figure}
	\centering 
	\includegraphics[scale=0.4]{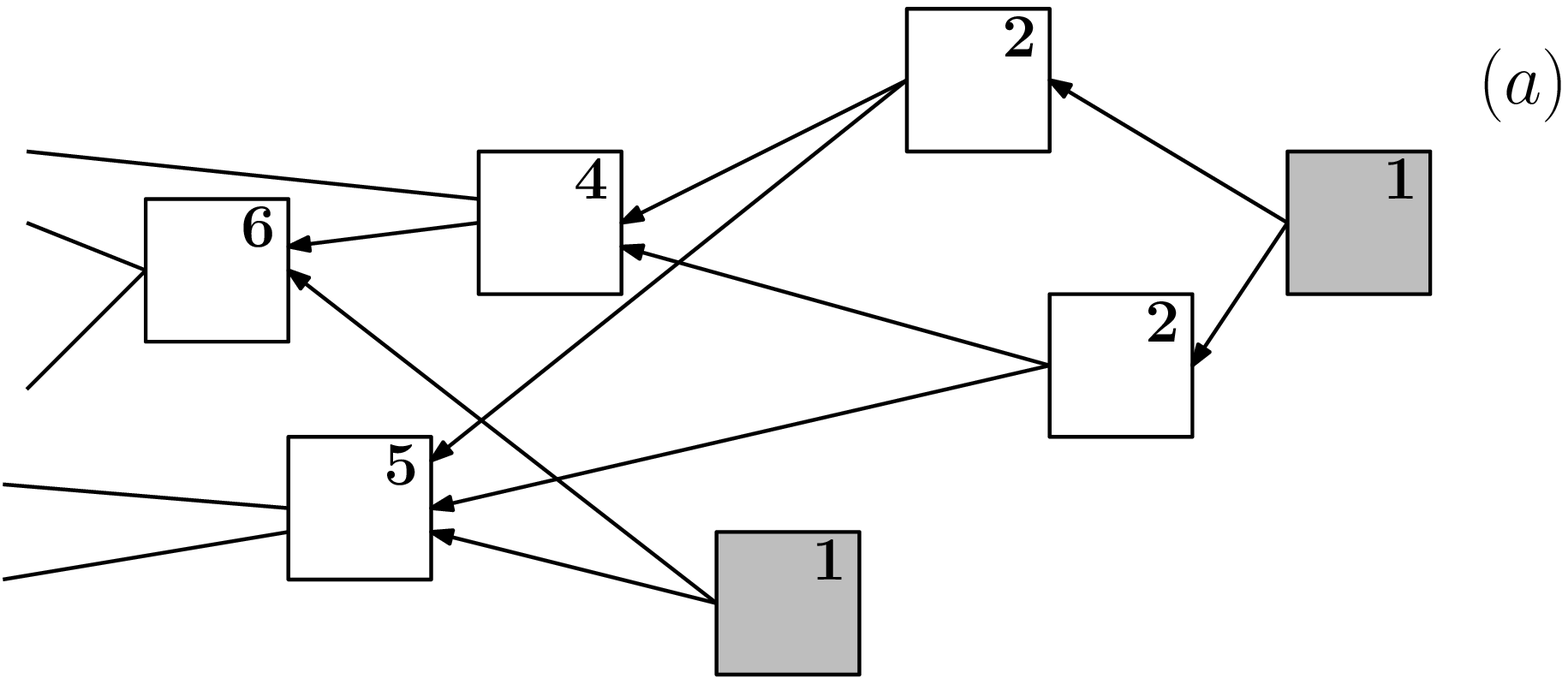}
	\includegraphics[scale=0.4]{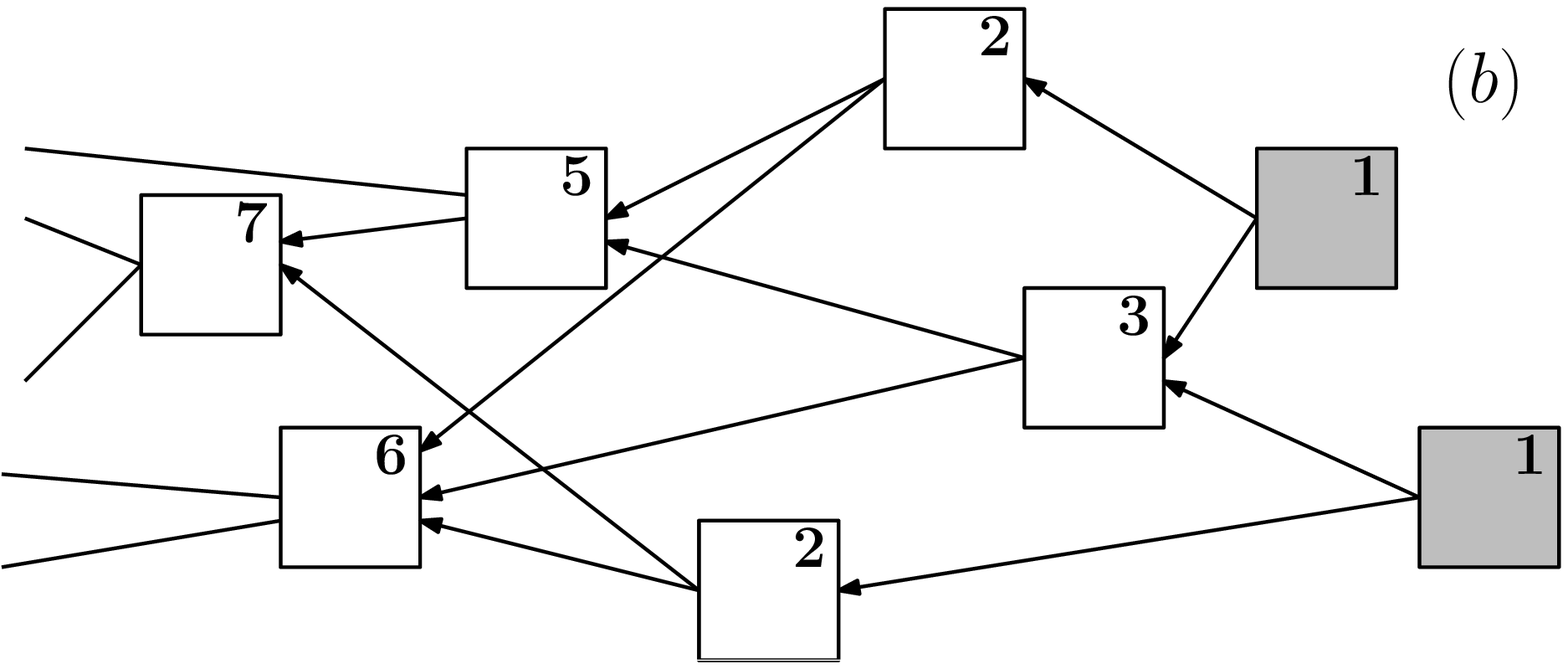}
	\caption{Examples of the Tangle. Transactions are represented by squares and approvals by arrows. Tips (unapproved transactions) are shown in gray. The cumulative weight of each transaction is in the  top right corner of each transaction. Parts $(a)$ and $(b)$ represent the Tangle before and after the new transaction has arrived respectively.   }\label{fig:Tangle_example1}	
	\end{figure}

 The Tangle \cite{I_WP,I_EQ} is a directed acyclic graph, growing over time. The graph begins with an initial transaction called the {\it genesis}.  As discussed in the introduction, each incoming transaction creates two edges to (i.e.\ approves) two previous,{\iffalse transactions with the IOTA protocol specifying that they should be\fi} nonconflicting tips. Thus, the Tangle never forms any cycles, as showed by Fig.\ \ref{fig:Tangle_example1}, and the genesis is always a minimum or terminal element.   
 
 In order to issue a previous transaction, a node must do {\it proof of work}, i.e. solve a cryptographic puzzle. A similar mechanism is used in Bitcoin to control the frequency of block creation.  The proof of work takes time, and hence there is a delay between the moment a transaction selects two tips and when the transaction appears in the Tangle. This delay increases while the transaction propagates throughout the network.  Let $h>0$ be the total length of this delay.  If $h=0$, the Tangle would be a chain, since there would only ever be one tip.  In this article, $h$ is a constant, although we could model it with a random variable.

The Tangle may fork due to conflicts.  For example, suppose a malicious actor double spends or makes two conflicting transactions $x_1$ and $x_2$.  Since new transactions cannot indirectly approve both $x_i$, two branches will form, with each branch consisting of transactions approving a single $x_i$.  

In this case, users must choose which branch is the ``true'' branch.  To do this, we introduce the {\it confidence level} of a transaction $x$ which is  the probability of the tip selection algorithm selecting a tip indirectly approving $x$.  The IOTA protocol dictates that only transactions or branches with sufficiently high confidence level should be trusted. This is akin to the ``6 block rule'' in Bitcoin. Thus the tip selection algorithm not only governs the growth of the Tangle, but is also the mechanism which achieves consensus.

Attackers can potentially cheat the system by manipulating the confidence level.  One method of doing this is called the ``parasite chain attack;'' see  \cite[Section 4.1]{I_WP}.  
The attacker issues a transaction $x$ in the main Tangle, paying a merchant for goods or services.  Meanwhile, the attacker issues transactions which are invisible to the public, forming what we call a {\it parasite chain}.  The parasite chain contains a  transaction $y$ which sends the money spent in $x$ to a different account.  After the confidence level of $x$ is high enough for the merchant to accept and deliver goods or services, the attacker then publishes the parasite chain. 

If successful, the tip selection algorithm will favor the tips of the parasite chain.  Thus new transactions will  approve the double-spend $y$ instead of $x$, and the parasite chain will grow at the expense of the main Tangle.  Meanwhile, the confidence level of $x$ decreases while the confidence level of $y$ becomes sufficiently high to be accepted by other users, thus invalidating $x$.

In Section  \ref{sec:safety}, we discuss how both URTS and URW are susceptible to parasite chain attacks and thus are unsatisfactory in the face of malicious actors. 
%
For these reasons, in its current implementation, the IOTA protocol uses neither of these tip selection algorithms and instead uses biased random walks\footnote{Currently, the protocol also uses a special node dubbed the {\it coordinator} whose role it is to prevent double spend attacks. Eventually, the coordinator will be removed in favour of a safe tip selection algorithm such as a BRW with a proper value of alpha.}.
However, URTS and URW still warrant attention for the following reasons.

\begin{enumerate}
    \item  The IOTA protocol is unable to enforce a particular tip selection algorithm: a user may choose any algorithm they like.  Researchers are thus behooved to understand a plethora of algorithms, in order to understand the choices available to users and also to find the best algorithms.   The IOTA protocol is relatively new and is not as extensively studied in published works. Thus simple algorithms like URTS and URW are a natural place for research to begin.

\item In the absence of a malicious actor, URTS and URW have many desirable properties.
In particular, URTS minimizes the time until first approval of transactions: see Subsection \ref{exprob}. Furthermore, URTS requires significantly less computational effort than a RW guided tip selection. 
Moreover, both URTS and URW do not leave orphans (valid transactions that never join the ledger). This means that the throughput is only constrained by the limits of the network. 
An optimal tip selection strategy should both provide safety and have these properties.

\item Both URTS and URW are idealized version of the Tangle and provide simplified scenarios and which are easy to analyze. 
Furthermore URTS allows for analytical derivations without considering the complex topological structure of the Tangle.
Similarly derivations in URW  disregard changes in cumulative weights\footnote{The problem is harder because cumulative weights change with time.} which are important in biased random walks. 
As a result, simulations of these algorithms have low overhead, see Section \ref{complexity}, and thus are easy to perform.
%

\item The biased random walk implemented in the IOTA protocol is guided by small value of $\alpha$ parameter (see section \ref{bias}).
For suitably small $\alpha$, the exit probabilities of the random walks will be similar to URW unless users attach tips to very old transactions.  Thus without malicious actors, the Tangle will grow similarly under both tip selection algorithms.

\end{enumerate}

As such URTS and URW are relevant models of the Tangle when all of the participants of the network are acting honestly. Other tip selection mechanisms that provide safety against attackers should be examined further, however this discussion is beyond the scope of this paper. 
 %
 %
%

Note that the URW is not equivalent to URTS. Any random walk on the Tangle is dependent on the topology of the DAG, while the URTS does not. 
In fact, these differences can be noticed in our simulations e.g.\ the average number of tips is higher with URW than URTS; see Section \ref{numberoftips} and \ref{exprob}.

\begin{figure*}[t]
     \centering
        \subfigure[URTS]{%
            \label{fig:cw_urst}
					\scalebox{0.65}{\input{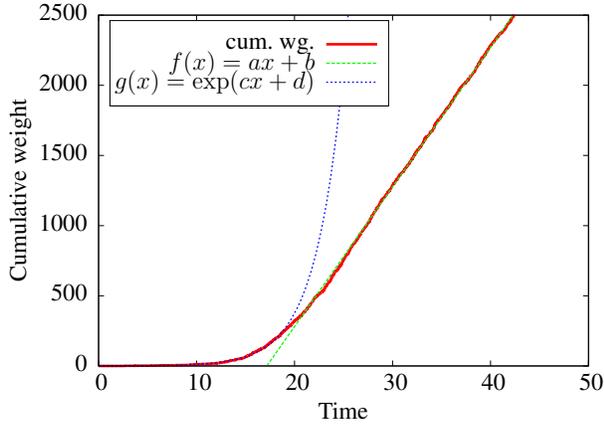}}
        }%
        \subfigure[URW]{%
          	\label{fig:cw_a=0}
					\scalebox{0.65}{\input{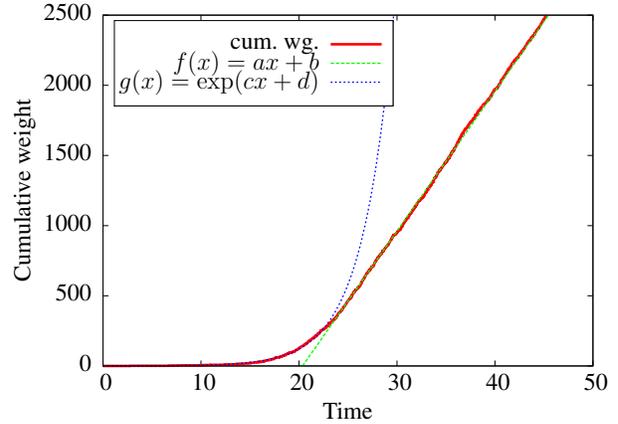}}
        } %
    \caption{%
        Cumulative weight of 200th issued transaction and the fitted linear and exponent functions ($f(x) = ax+b$, $g(x) = \exp(cx+d)$ respectively). $\lambda = 100$.
     }%
   \label{cw}
\end{figure*}

\section{Results}\label{results}

\begin{figure}[h!]
\centering
\scalebox{0.65}{\input{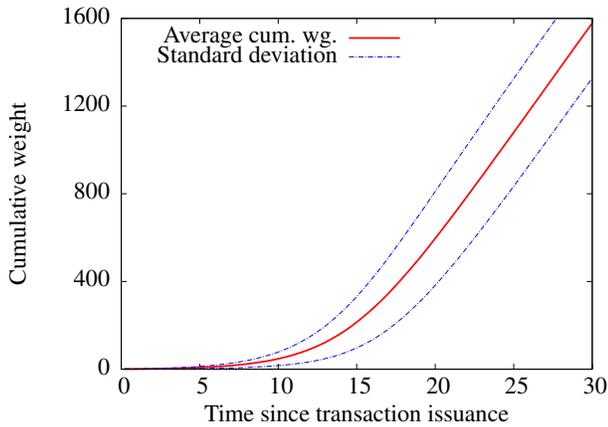}}

\caption{Average cumulative weight with the time since a transaction was issued. Tip selection is URW and  $\lambda=100$.}.
\label{fig:cw_many}
\end{figure}

In this section, we present the results of our simulations. We begin with analyzing the growth of cumulative weight and then discuss the number of tips,  $L(t)$ as a function of time, and also the time until the first approval. 
Lastly, we use exit probabilities to explain observed differences in behaviours of URTS and URW Tangles, and we also present the growth of the simulations' overhead as a function of txs in the Tangle. 

The arrival of new transactions is modeled through the standard approach of a Poisson point process \cite{I_WP}.
We denote the rate of this process by $\lambda$ and assume that this rate remains constant in time. Then the elapsed time between two consecutive transactions is in turn given by the exponential distribution $\mathbf{Exp}(\lambda)$. 

Recall that $h$ is the delay between a transaction selecting tips and being added to the Tangle.  Since $h$ is measured in units of time, and $\lambda$ in transactions per unit of time, the properties of the Tangle only depend on $\lambda\cdot h$.  Thus, we can fix our time scale by setting $h=1$, or equivalently we consider  time measured in units $h$. 
For simplicity,  we omit the $h$ throughout the remainder of the paper.

In most cases, we focus on $\lambda = 100$, but some results are presented with $\lambda$ ranging up to $10^4$. We chose these numbers because they are the approximate orders of magnitude of the protocol in high load regime.  
Simulations were performed for Tangles with $10^{5}-10^7$ transactions. 
We focus on the steady state of the Tangle, thus, presented data are recorded after an initiation phase of $50h$ from the Genesis.
Some of the presented data (Figs. \ref{cw}, \ref{ev} and \ref{fig:tips_alpha}) are included to illustrate the evolution of a single Tangle, however the remaining results and figures are obtained by averaging over more than $10^3$ simulation samples. Our results have a significance level of $0.01$, corresponding to a confidence level of 99\% with a width of the confidence interval of, at most, $0.1\%$ of the respective mean value.

\subsection{Cumulative weight}

Supporting findings in \cite[Section 3.1]{I_WP}, the data suggests that initially the cumulative weight of a transaction grows exponentially during what is called the  {\it adaptation period}, and then grows linearly with slope $\lambda$; see Figs.\ \ref{fig:cw_urst} and \ref{fig:cw_a=0}.   This is the case for both examined tip selection algorithms, URTS and URW. Fig.\ \ref{fig:cw_many} shows that this behaviour is typical of transactions under URW.  

These results are intuitive.  Consider a new transaction $x$.  Initially, $x$ will be approved by several new transactions, and those transactions will again be approved by several transactions.  Thus the cumulative weight of $x$ will initially grow exponentially during the adaptation phase.  However, eventually  most tips will indirectly approve $x$. At that point, most incoming transactions will also indirectly approve $x$ and will contribute to its cumulative weight.  So eventually the cumulative weight of $x$ will grow at the same rate as the Tangle, i.e.\ linearly with slope $\lambda$.

Because our data is discrete and not continuous, the transition point between these growth phases is difficult to rigorously determine. However, by the previous discussion, we know that it occurs when the confidence level is close to $100\%$.

\subsection{Number of tips} \label{numberoftips}

\begin{table}[h!]
\caption{ Average number of tips and its standard deviation $\sigma$ as a function of $\lambda$.}
\begin{center}
{\footnotesize
\renewcommand{\arraystretch}{1.25}%
\begin{tabular}{|c||l|l||l|l|}
\hline
 \multirow{ 2}{*}{ $\lambda$ }& \multicolumn{2}{c||}{URTS }       &  \multicolumn{2}{c|}{ URW  }     \\\cline{2-5}
	&	 $\mathbb{E}L$ 	& $\sigma$ 	& $\mathbb{E}L$ 	& $\sigma$ \\
\hline
   1   	& 2.858   &    1.115  &  3.035   &    1.167    \\\hline
   10	& 20.69   &    3.140  &  21.80   &    3.278    \\\hline
   100	& 200.9   &    10.21  &  208.1   &    10.27    \\\hline
   1000	& 1999    &    28.57  &  2079    &    34.36    \\\hline
   10000& 19983   &    86.28  &  20722   &    90.41    \\\hline

\end{tabular}
\label{tab:average}}
\end{center} 
\end{table}

\begin{figure*}
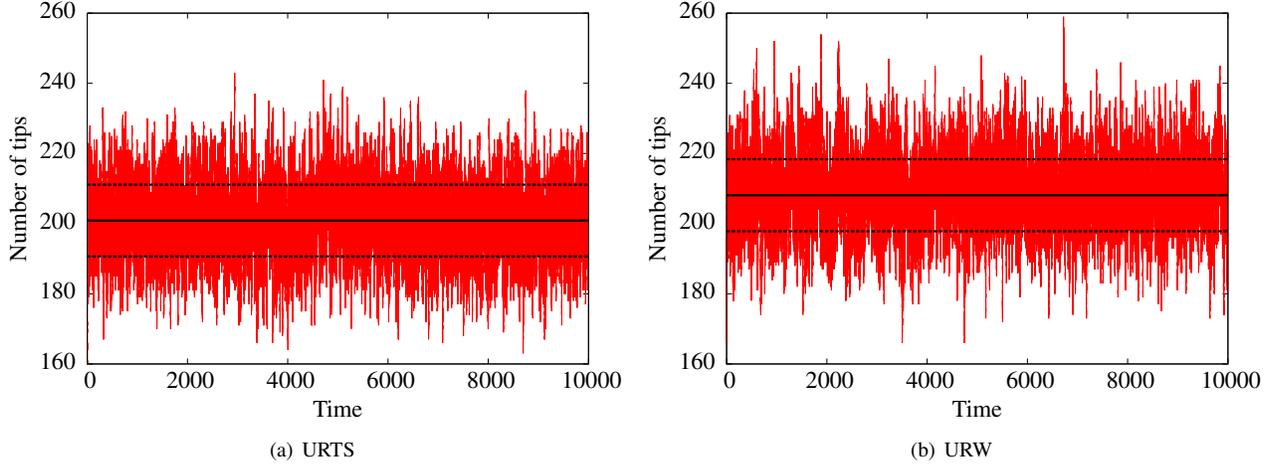

     \centering
        \subfigure[URTS]{%
            \label{tp_URST}
					\scalebox{0.65}{\input{ev_urst.tex}}
        }%
        \subfigure[URW]{%
          	\label{tp_MCMC}
					\scalebox{0.65}{\input{ev.tex}}
        } %
    \caption{%
        Evolution of $L(t)$, the number of tips at time $t$.  $\lambda = 100$.
     }%
   \label{ev}
\end{figure*}

\begin{figure}
\begin{center}    
 \scalebox{0.650}{\input{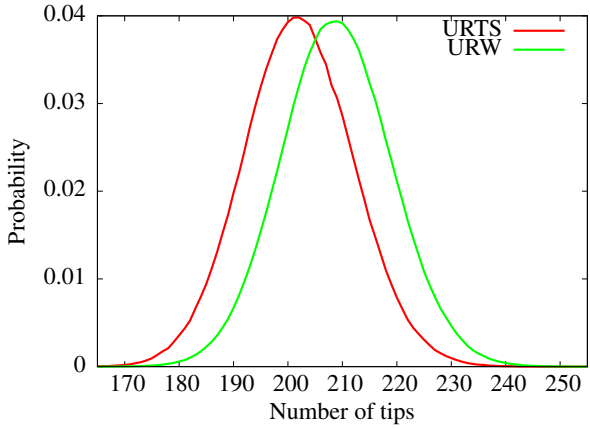} }        
  \caption{ Empirical probability density function of  the number of tips, $L(t)$, (normalized histogram) with  $\lambda = 100$.}\label{fig:his} 
  \end{center}  
\end{figure}

The typical evolution of the number of tips, $L(t)$, for URTS and URW is presented in Figs.\ \ref{tp_URST} and \ref{tp_MCMC} respectively. 
In both cases, $L(t)$ is bounded and fluctuates around an average value. 
The dashed lines in the plots represent the standard deviation of the data. 
From these plots, we empirically construct the probability density functions of  $L(t)$; see Fig.\ \ref{fig:his}.

\begin{figure} 
\begin{center}    
 \scalebox{0.650}{\input{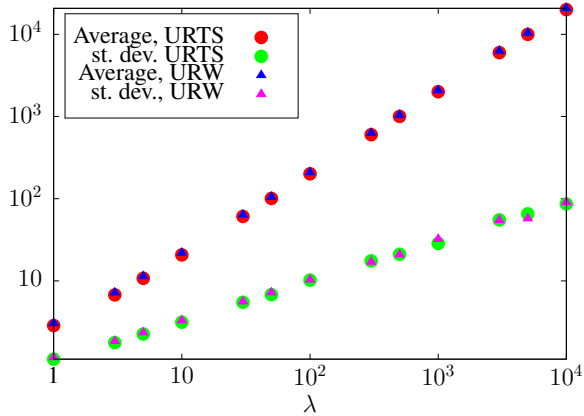} }        
  \caption{ Average number of tips $\mathbb{E}L$ (and its standard deviation) as a function of $\lambda$ in log-log scale. It illustrates that the average (standard deviation) scales linearly (as a square root) with the value of $\lambda$. }\label{fig:scaling} 
  \end{center}  
\end{figure}

The empirical probability density functions are bell curves and are presented in Fig. \ref{fig:his}.  The curve for URW shifted to the right of URTS. Note that these histograms are not fitted.  Although not recorded here, simulations with other values of $\lambda$ yielded similar curves. The means and standard deviations resulting from these simulations are displayed in Table \ref{tab:average} and Fig. \ref{fig:scaling}. 

The results in Table \ref{tab:average} confirm the analytic predictions \cite[Section 3 Equation (1)]{I_WP} that the mean tip number 
for URTS is $2\lambda$. For URW, the average value of $L(t)$  is approximately $2.1 \lambda$ which is $3.5-4.5 \%$ higher than under URTS. Furthermore, the standard deviation of the empirical probability density of $L(t)$ is proportional to the square root of $\lambda$; see Fig.\ \ref{fig:scaling}.
The difference between the number of tips between URTS and URW is interesting, since it demonstrates that the Tangle grows differently with these tip selection algorithms. We explore the reasons for the differences in Subsection \ref{exprob}.

\subsection{Time until the first approval}

The average times until the first approval for URTS and URW are presented in Table \ref{tab:average_time}. The data demonstrates that URTS provides slightly shorter times than URW (approximately $2$ verses $2.1$). Furthermore, the standard deviation of this average is significantly smaller for URTS.  Moreover, our data agrees with the prediction made in the white paper for URTS \cite[Section 3]{I_WP}.

The average time till approval is actually related to the number of tips.  Indeed, consider the following derivation in the white paper \cite[Section 3]{I_WP} for the average time approval under URTS.  Since $\lambda$ is the rate of incoming transactions, and each transaction approves two tips, all the tips combined receive approvals at a rate of $2\lambda$.  Since a tip will be selected with probability $1/L(t)$ under URTS, the rate, $R$, that it receives approvers should be
\begin{equation}\label{relation2}
R=\frac{2\lambda}{L(t)}.
\end{equation}
Let $t_A$ denote the average time till approval. Since $t_A$ includes the $1h$ time delay,  we have $R=1/(t_A-1)$ and thus
\begin{equation}\label{relation}
L(t)=(t_A-1)2\lambda.
\end{equation}
When $L(t)=2\lambda$, we derive $t_A=2$.

This derivation does not follow for URW, or any other tip selection algorithm, since the probability of each tip being selected is not equal.  In fact, we can see that the data in Tables \ref{tab:average} and \ref{tab:average_time} does not support the relationship in  Equation \ref{relation} under URW.  However, interestingly the values are not completely dissimilar. It would be interesting to derive a general formula relating $L(t)$, $\lambda$, and $t_A$.

\begin{table}[h!]
\caption{ Average time until the first approval and its standard deviation $\sigma$ as a function of $\lambda$.}
\begin{center}
{\footnotesize
\renewcommand{\arraystretch}{1.25}%
\begin{tabular}{|c||l|l||l|l|}
\hline
 \multirow{ 2}{*}{ $\lambda$ }& \multicolumn{2}{c||}{URTS }       &  \multicolumn{2}{c|}{ URW  }     \\\cline{2-5}
	&   $t_A$    &   $\sigma $    &   $t_A$    &   $\sigma$ \\
\hline
   1   	& 2.884 &   1.951  &  3.042  &  2.464    \\\hline
   10	& 2.075 &   1.082  &  2.177  &  1.482    \\\hline
   100	& 2.009 &   1.011  &  2.087  &  1.349    \\\hline
   1000	& 1.997 &   0.997  &  2.078  &  1.345    \\\hline
\end{tabular}
\label{tab:average_time}}
\end{center} 
\end{table}

%
%
%
%
%
%
%
%
%

For both considered tip selection algorithms, URTS and URW, since the number of tips is constant every transaction will eventually be approved almost surely. However, this may not be the case with other tip selection algorithms. For example simulations in \cite{POBLB} reveal that while using the tip selection algorithm discussed in Section \ref{bias},     some transactions will always remain tips, i.e.\ be {\it permanent tips}.  Permanent tips make computing the average time until approval problematic.  Since their approval time is infinite, they cannot be included in the average, but excluding them may artificially decrease the average and  lead to false conclusions.

\subsection{Biased Random Walks}\label{bias}

The IOTA protocol uses a random walk biased towards transactions with large cumulative weight \cite[Section 4.1]{I_WP}. The amount of bias is determined by a parameter $\alpha\ge 0$.  Specifically, the probability $P_{xy}$ of transitioning\footnote{Recall that our walks move backwards through the Tangle, moving from approvee to approver.} from $x$ to $y$ with $y\to x$ when $y$ approves $x$ is
\begin{equation}\label{eq:trprob}
P_{xy} = \frac{\exp \left(\alpha\mathcal{H}_y  \right)}{  \sum_{z:z\to  x} \exp \left (\alpha\mathcal{H}_z \right ) }  
\end{equation}
where $\mathcal{H}_z$ denotes the cumulative weight of the transaction $z$ and the sum is taken over all transactions $z$ which approve $x$. We use BRW to denote the tip selection algorithm using these biased random walks beginning at the genesis. BRW with $\alpha=0$ is simply URW. We note that both URW and BRW are special cases of the {\it Monte Carlo Markov chain} algorithms proposed in \cite{I_WP}.

The transition probability in (\ref{eq:trprob}) is similar to the Gibbs distribution from physics  \cite{LL_statistical}, where cumulative weight plays the role of the energy\footnote{Or actually negative energy (energy multiplied by $-1$), since the walks gravitate to higher cumulative weights, just like thermodynamical systems tend to favor states of smaller energy.} of state (or transaction in our case) and $\alpha$ is the inverse temperature of the system. When $\alpha$ is large, the system is ``cold''  and random walks move only along a relatively small number of paths with the highest weight transactions. On the other hand, when $\alpha$ is small, the system is ``hot''  causing particles to move more chaotically.

\begin{figure} 
\begin{center}    
 \scalebox{0.650}{\input{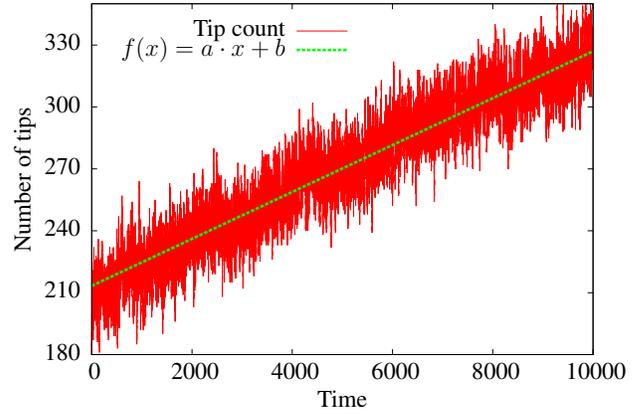} }        
  \caption{ Evolution of $L(t)$, the number of tips at time $t$. $\lambda = 100$; Biased random walk with $\alpha = 0.003$. The growth of the tip number is linear and for illustration fitted by  $f(x)= a \cdot x +b$.  }\label{fig:tips_alpha} 
  \end{center}  
\end{figure}

Fig. \ref{fig:tips_alpha} shows that the number of tips grows linearly  even for small values of $\alpha>0$. When $\alpha$ is large, paths of random walks gravitate towards paths with the highest cumulative weight, increasing the likelihood that certain tips would be approved. These tips then gain more approvers, their cumulative weight grows, and then they again pull the random walks in their direction.  Meanwhile, other tips are becoming increasingly less likely to receive approvers and remain tips longer, even as fresh tips still continue to arrive.  Thus the tip number increases, explaining the trend in Fig. \ref{fig:tips_alpha}. 
{
Transactions whose cumulative weight growth has stalled, i.e. they have not been directly or indirectly approved for an excessive amount of time, are considered orphaned. Since their confidence level tends to zero these transactions are not considered as part of the consensus.
}

\subsection{Exit probabilities}\label{exprob}

For a vector of available tips $\overline{tp} =(tp_1,tp_2,...)$, a tip selection algorithm determines a corresponding vector of exit probabilities $\overline{p} = (p_1,p_2,...)$ where $p_i$ is the probability that the tip $tp_i$ will be selected.  We call these values {\it exit probabilities}.
%
%
Since $\sum_i p_i = 1$, the exit probabilities define a probability space on the set of tips.

For any algorithm using a random walk, the corresponding exit probabilities will generally not be equal because the topology of the underlying graph is not uniform.

\begin{figure} 
\begin{center}    
 \scalebox{0.650}{\input{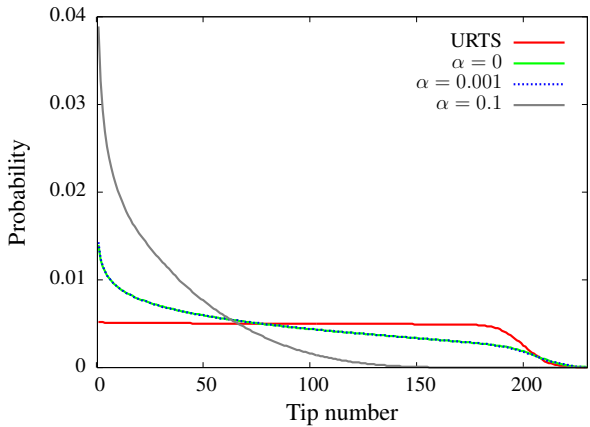} }        
  \caption{Comparison of the adjusted average exit probability of the $i$-th most probable tip, for $\lambda=100$ and different tip selection algorithms: URTS, URW and BRWs ($\alpha = 0.001, 0.1$). }\label{figexpr_com} 
  \end{center}  
\end{figure}

We illustrated the phenomenon in Fig. \ref{figexpr_com}, where we present what we dub the {\it adjusted average exit probability of the $i$-th most probable tip} for a Tangle with $\lambda=100$ and for different tip selection algorithms.
The plot is constructed in the following way. We simulate $10^3$ Tangles. For each of these samples we calculate the exit probability distribution on the last set of tips by running the tip selection algorithm $10^6$ times.  We then order these measured exit probabilities from greatest to smallest. Since the number of tips will vary between the different samples, for each sample, we extend with zeroes the ordered exit probability vector $\bar p$ to the maximum observed tip number. 
Finally we average these vectors over all simulation runs. Hence, due to the variation in the tip number, the curves show a decaying trend for the higher indices. 


The more the exit probabilities between individual tips vary, the steeper the slope of the curve in Fig. \ref{figexpr_com}. Because of outliers in the sampling and varying tip numbers, the curve is not flat even for URTS.  In Fig. \ref{figexpr_com}, the curves for $\alpha = 0$ and $\alpha = 0.001$ closely overlap whereas the curve for $\alpha =0.1$ differs. 
This suggests that when $\alpha = 0$ and $\alpha = 0.001$ the Tangle should evolve similarly because the exit probabilities determine the growth.
 Currently $\alpha =0.001$ is used in the IOTA protocol, specifically in the IOTA Reference Implementation; see IOTA Git Hub repository \cite[line 816]{github_iotanode}.
The similar behaviour of Tangles with URW and this value of $\alpha$ used in  IOTA Reference Implementation justifies our focus in this article on URW: see Section \ref{Tangle}, point 4).

%
%
%

%
%
%
%

As we previously demonstrated, the average tip number and the time till first approval is increased for URW compared to URTS. 
This is associated with the different exit probabilities on the tips, since the average flow rate of direct approvals of a tip $tp_i$ is proportional to the probability $p_i$. 
Therefore, when the $p_i$ are not equal, some transactions will receive more direct approvals, and 
other tips will wait much longer for their first approval.
In fact, this is why we observe that URTS provides on average a shorter time until first approval. 
%
This is because exit probabilities for URTS are maximally spread out.

\subsection{Performance evaluation}\label{complexity}
\begin{figure} 
\begin{center}    
 \scalebox{0.650}{\input{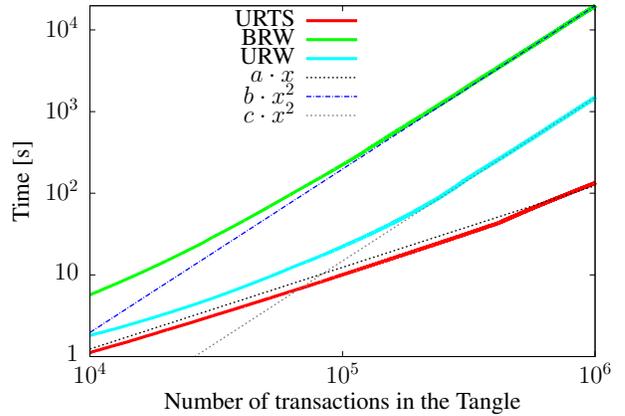} }        
  \caption{ Time required to complete simulations of Tangles of different size. The computational overhead for URTS grows linearly (linear fit with black dashed line), whereas with URW and BRW the overhead grows proportionally to the square of the total number of transactions (square function fit, blue and gray dashed lines).  }\label{fig:compl} 
  \end{center}  
\end{figure}

In this section we compare how tip selection algorithms affect performance in terms  of the computational overhead.
This information is relevant for the actual implementation of Tangle based DLTs because the simulation complexity is similar to the actual protocol implementation performance complexity. 
Preferably, tip selection mechanism should be as efficient as possible, 
especially if the considered protocol wants to accommodate many transactions per second, such as in platforms created for IoT markets.
We run simulations varying the total number of transactions, and we measure the time required to complete the simulation using a) URTS; b) BRW with cumulative weight update; c) URW without a cumulative weight update.  Fig. \ref{fig:compl} and Table \ref{tab:time_complexity} summarize our results.
The red curve corresponds to URTS based simulation. 
The green and cyan curves correspond to the BRW and URW based simulation with and without an update of cumulative weight respectively, and we fit times with functions a) $a\cdot x$; b) $b\cdot x^2$; c) $c\cdot x^2$ respectively. 
%

%
Simulation performance when using URTS only depends on the number of tips, which is bounded and on average equal to $\approx2\lambda$. 
Since we can disregard the cumulative weights in this setting, the computational overhead grows linearly with the number of transactions in the Tangle, thus with time complexity of order $\mathcal{O}(n)$. 

On the other hand, the performance of simulations based on URW depends on the length of the random walk, which increases linearly with the Tangle size, leading to a square relationship between the computational overhead and the number of transactions in the Tangle with a time complexity of order $\mathcal{O}(n^2)$.

Similarly, BRW depends on the length of the random walk as well as on the cumulative weight of all the transactions, and so as new transactions enter the Tangle, the cumulative weights of all approved transactions need to be updated. Thus, when cumulative weights are updated using {\it depth first search} of the Tangle, the time needed to run simulations grows like the square of the number of transactions in the Tangle, resulting in a time complexity of order $\mathcal{O}(n^2)$. 

Overall, both BRW and URW curves grow quadratically with the number of transactions, although simulations with URW are less demanding.

\begin{table}[h!]
\caption{Computational overhead of simulations based on URTS, BRW and URW tip selection algorithms.}
\begin{center}
{\footnotesize
\renewcommand{\arraystretch}{1.25}%
\begin{tabular}{|c|c|}
\hline
Tip selection algorithm & Time complexity	\\\hline

 URTS 	&     $\mathcal{O}(n)$     			\\\hline
 BRW with cumulative weight update  	&    $\mathcal{O}(n^2)$   \\\hline
 URW without cumulative weight update  	&    $\mathcal{O}(n^2)$   \\\hline
\end{tabular}
\label{tab:time_complexity}}
\end{center} 
\end{table}

\section{Tip selection mechanism and safety of the network}\label{sec:safety}
Both URTS and URW are theoretical tip selection algorithms and should be treated merely as a tool to study the Tangle. 
As discussed in Section \ref{Tangle}, neither of them can be used in a real-life implementation of DAG based DLT, because they are not safe against parasite chain attacks. In this section, we illustrate why.  The arguments in this section are known to experts on the Tangle, however they have never been published.

\begin{figure}
	\centering 
	\includegraphics[scale=0.6]{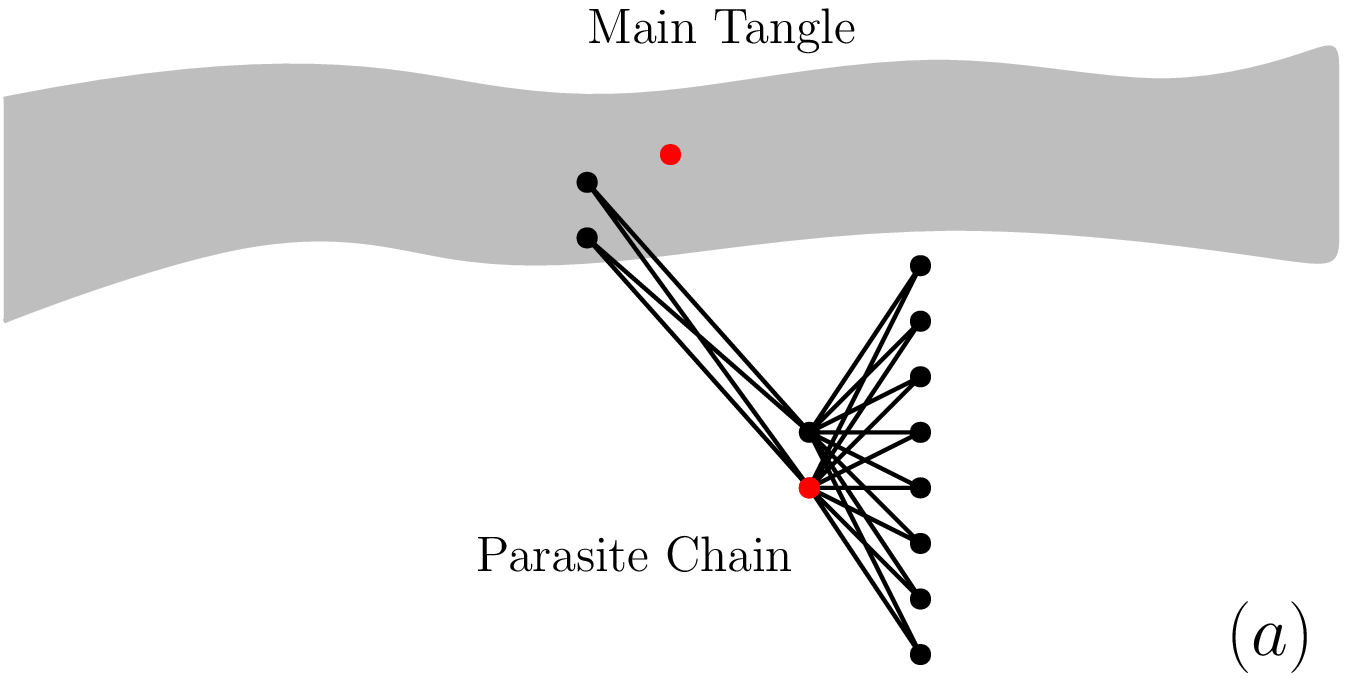}\\
	\vspace*{12mm}
	\includegraphics[scale=0.6]{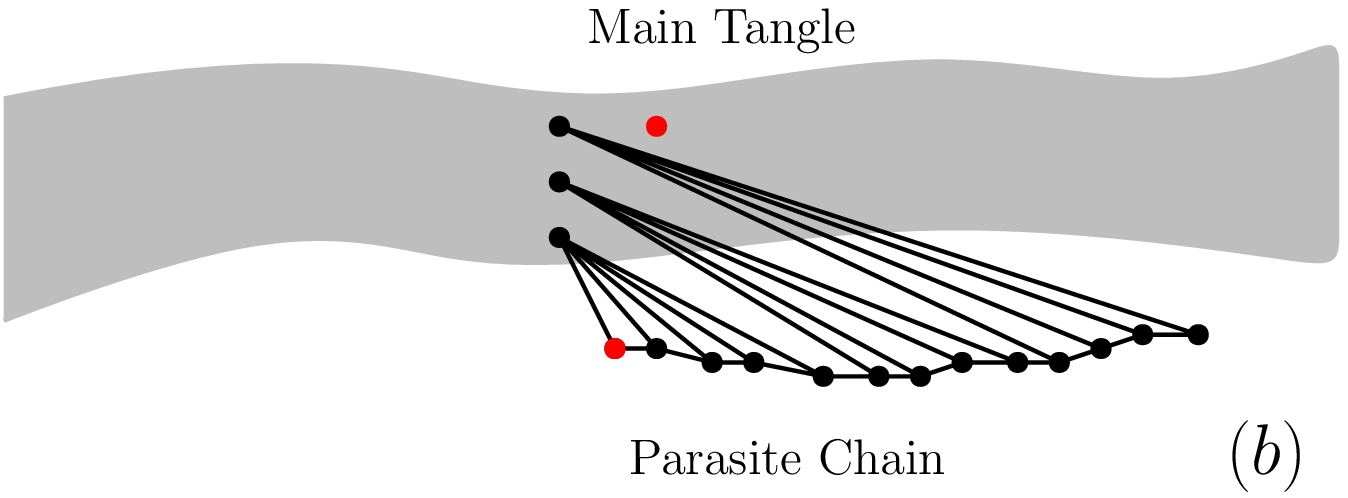}
	\caption{The examples of parasite chains attacks which will be successful against URTS (part $(a)$) and URW (part $(b)$). Conflicting transactions are in red.}\label{fig:PC}	
	\end{figure}

In the case of URTS, the attacker proceeds in the following manner. 
After placing the first double spend transaction in the main part of the Tangle, the attacker secretly issues the second double spend transaction.  The rest of the parasite chain consists of tips approving the second double spend and some other transaction; see Fig. \ref{fig:PC} part $(a)$.
The attacker publishes  the parasite chain as soon as it has significantly more tips than the main Tangle.  
Then most of the honest users will approve the second double spend transaction.

Both the white paper \cite[Section 3 Equation (1)]{I_WP} and our simulations (see Section \ref{numberoftips}) show that the number of tips in the main Tangle is stable and is around $2\lambda$. Thus the attacker is guaranteed to be able to produce more parasitic tips.

URW provides protection against this type of parasite chain, (see part $(a)$ of the Fig. \ref{fig:PC}) because URW is unaffected by a number of tips in each of the sub-Tangles.

However, URW can be attacked with the following parasite chain displayed in Fig. \ref{fig:PC} part $(b)$.
First, the attacker chooses a set $Y$ of transactions in the main Tangle such that every path from the genesis to a tip passes through an element in $Y$.  Then, in secret, the attacker issues a transaction $x_0$, and then issues a sequence of transactions $x_1,x_2,x_3,\dots$ such that each $x_i$ directly approves $x_{i-1}$ and some transaction in Y.  

Suppose that transactions are accepted once their confidence level is greater than some $\kappa\in (0,1)$.  The attacker continues until for each $y\in Y$, the proportion  of direct approvers of $y$ in the parasite chain is greater than $\kappa$.  Then they issue a transaction $y$ double spending $x$.  After the confidence level of $y$ is greater than $\kappa$,  the attacker publishes the parasite chain.
At this point, the probability of a random walk ending on the parasite chain is greater than $\kappa$.   Moreover once on the parasite chain, the random walk cannot move out of it and must approve $x_0$.  Thus the confidence level of $x_0$ is greater than $\kappa$, and the attack is successful.

The parasite chains discussed in this section however are foiled by the biased random walks discussed in  Section \ref{bias}.  Indeed, biased random walks are less influenced by small changes in the Tangle, and thus an attacker must use a large amount of hashing power to effectively attempt an attack.

A BRW with higher $\alpha$ provides more safety.  Indeed,  the main tangle will always be heavier than any parasite chain, assuming of course that the attacker does not control a majority of the hashing power.  Thus, the larger $\alpha$ is,  the more biased random walks are inclined to stay on the main tangle and eschew any parasite chains.  
%
%
{
However, a large $\alpha$ value also leads to a higher orphanage rate of transactions, see Section \ref{bias}. Hence an optimal selection of the value of $\alpha$ should ensure both the safety as well as the liveness of the protocol. 
}
%

\section{Conclusions}\label{sec:disc_conslusion}

In this paper, we introduced and discussed several of the basic concepts and properties of the Tangle, such as tips, tip selection algorithms, and cumulative weight. 
We focused on the Uniform Random Tip Selection (URTS) which selects tips with uniform probability from the set of available tips, and the Unbiased Random Walk (URW) which uses a random walk beginning at the genesis. 
%
%
Our results confirms that cumulative weight grows in two phases: an initial phase of exponential growth, followed by a linear phase during which the transaction is indirectly approved by effectively all new transactions. 
We found that the average value of tips agrees well with the theoretically predicted value for URTS and hence our data justifies the approximations used in the derivation of these predictions.  
%
Moreover our simulations reveal that the average number of tips is higher with URW than URTS and that URW increases the time till first approval by a similar ratio.
We than explain this behaviour by providing the numerically obtained exit probabilities, which are hard to calculate analytically. 
We also considered some biased random walks (BRW) and showed their relationship with permanent tips.

Finally, we discussed and compared the computational overhead of our simulations when using URTS, URW, and BRW. 
Our results show that time complexity of simulations based on URTS grows linearly with the size of the Tangle and are significantly more efficient than simulations based on both URW and BRW, whereas their time complexity exhibit quadratic growth.

This information is relevant in implementation of nodes operating DLTs based on the Tangle. 
We believe these results can serve as a fundamental ground to analyze and improve the\bart{\iffalse Tangle based protocols as well as to foster \fi} research and development\bart{\iffalse aimed at improving\fi} of DAG-based DLTs. 
In particular, to compute the IOTA throughput we must know the probability that the confidence level of a transaction will tend to 0, which is related to the evolution of the number of tips.  
We also would like to understand the time it takes for the confidence level to approach 1, a number dependent on the time till first approval. 
Lastly, the security of the Tangle depends on the inability of an attacker to manipulate the confidence level of transactions.  
However, with the biased random walks, the confidence level is essentially determined by the cumulative weights.  
%

\balance

\bibliographystyle{IEEEtran}
\bibliography{bibliography}

\end{document}